# Physical Proximity and Spreading in Dynamic Social Networks

Arkadiusz Stopczynski[1,2], Alex 'Sandy' Pentland[2], Sune Lehmann[1,3]

[1]*Department of Applied Mathematics and Computer Science, Technical University of Denmark, Kgs. Lyngby, Denmark*

[2]*Media Lab, Massachusetts Institute of Technology, Cambridge, MA, USA*

[3]*The Niels Bohr Institute, University of Copenhagen, Copenhagen, Denmark*

**Most infectious diseases spread on a dynamic network of human interactions. Recent studies of social dynamics have provided evidence that spreading patterns may depend strongly on detailed micro-dynamics of the social system. We have recorded every single interaction within a large population, mapping out—for the first time at scale—the complete proximity network for a densely-connected system. Here we show the striking impact of interaction-distance on the network structure and dynamics of spreading processes. We create networks supporting close (*intimate network*, up to $\sim 1$ m) and longer distance (*ambient network*, up to $\sim 10$ m) modes of transmission. The intimate network is fragmented, with weak ties bridging densely-connected neighborhoods, whereas the ambient network supports spread driven by random contacts between strangers. While there is no trivial mapping from the micro-dynamics of proximity networks to empirical epidemics, these networks provide a telling approximation of droplet and airborne modes of pathogen spreading. The dramatic difference in outbreak dynamics has implications for public policy and methodology of data collection and modeling.**

Today, even after decades of effort, we are only beginning to understand the detailed process of how epidemics spread through populations[1–8]. As large-scale data on human behavior has grown abundant, we have seen dramatic improvements in epidemic prediction driven by access to large-scale travel patterns[9–12], while data on empirical communication patterns of millions of phone subscribers have enabled an understanding of the complex relationship between network structure and link weights[13]. Starting from a unique dataset[14] detailing every proximity event in a large population of around $500$ individuals, we explore the dynamic network where every person is represented by a node, and two nodes are connected if they are within certain distance $d$ of each other. Specifically we focus on spreading in two distinct physical proximity networks. We consider the network of *intimate* interactions defined as $d \lesssim 1$ meter, and the *ambient network* which includes all interactions $d \lesssim 10$ meters[15] (see also Methods and SI). Below we show that the intimate and ambient networks are fundamentally different in terms of structure and dynamics. Using a simple model of epidemic spreading on these empirical networks of interactions, we explore the implications for disease spreading and find that the network properties strongly influence spreading outcomes, emphasizing the differences between the two networks.

In the literature, the tacit rules of human interactions in physical space have been an object of interest since the 1950's[16–18]. Yet, little is known about the fundamental structure of potential spreading networks created by different levels of interactions proximity, the dynamics of spreading supported by such networks, or the properties of resulting outbreaks. While previous research into proximity networks has relied on self-reported data[17,19] or tightly-controlled laboratory observation[16], we focus on empirical large-scale data collected using modern smartphones[1,14,18,20]. Recent studies on small-scale high-resolution human interaction data have hinted that the local infection dynamics are crucially dependent on the detailed micro-dynamics of interactions in the social system[1,7,21–23].

Physical proximity interactions provide a fundamental mode of transmission for pathogen spreading[1,4,19], and the two most prominent modes of pathogen spreading are droplet- and airborne transmission[2]. These modes of spreading are driven by the size of droplets produced by the pathogen[24]. In the droplet case, large particles containing pathogen are created when sneezing or talking and need to directly land on a new host, otherwise quickly falling on the ground. This process has an effective range of transmission up to approximately $1$ meter[2]. The particles in the airborne mode of transmission are smaller ($< 5\mu$m) and can remain suspended in the air for long periods of time. Therefore, airborne transmission has a significantly increased range compared to droplet spreading, infection can occur for hours after an infected has occupied a room, and the rate of airborne transmission depends strongly on the environment.

While access to comprehensive empirical network dynamics represents a major step forward, simulating epidemics on the intimate and ambient network provides only a rough sketch of droplet and airborne spreading processes in real networks. The simulated spreading does not account for many details of real spreading[2,25], and particularly the ambient network does not account for important properties of airborne spreading, such as droplets suspended in the air for extended periods of time. Since the ambient network does not include this feature, the interactions on this network can be considered a subset of the possible paths supporting airborne spreading. As a consequence, we find that the difference between simulated spreading on empirical proximity networks is likely to be smaller than the difference between actual droplet and airborne disease spreading on the same population. Here we show that simply by increasing the range of interactions, simulated spreading processes on empirical contact networks behaves in a fundamentally different way. Our findings therefore reveal that droplet and airborne modes should be modeled in different ways and based on data corresponding to the disease transmission mode under study.

We expect the social network of individuals to be closely related to the structure of the intimate network, but with important differences. This connection arises because, in social networks, the difference between friend and stranger is typically expressed via different personal spaces for each social category[17]. Similarly, interactions with individuals with whom we are not familiar tend to occur at larger distances (we use term 'interaction' for all proximity events, including the ambient network). Since people function in bounded spaces, they do not have complete freedom to only allow friends to be physically close to them and to keep strangers further away. Rides on buses, random meetings in elevators, or busy cafeterias force us to be in close proximty to strangers. Thus, while the majority of our intimate interactions are with friends and families, our contact network is not fully explained by the underlying social network, as expressed by, for example, link strengths. This implies that considering the proximity of the interactions—going beyond the strength of the links in the network—provides a new source of information regarding potential spreading paths.

## Results

Our results are based on close proximity interactions between $N = 464$ highly-connected participants—freshmen students at a large university[14]. The full physical proximity network is created by interactions occurring at any distance covered by Bluetooth range, between $0$ and $10 - 15$ meters (ambient network). This densely-connected dynamic network of all Bluetooth interactions arises from a total of $1\,472\,090$ interactions, taking place over 28 days. We define an *interaction* between users $i, j$ in 5-minute timebin $t$ as $\gamma_{ijt} = s$, where the signal strength $s$ is reported by the handsets as received signal strength indicator (RSSI). RSSI, measured in $dBm$, is defined as the observed signal power relative to 1 $mW$. In order to capture close range interactions, we establish the intimate network by selecting the subset of interactions with $\gamma_{ijt} \geq -75\ dBm$ corresponding to distances of approximately $1$ meter or less[15]. This network consists of $f = 18.3\%$ of all interactions.

**Sampled ambient network.** Because the intimate network contains only a fraction of all interactions, spreading processes taking place on the this network are trivially slower and smaller than in the ambient case. This implies that we cannot directly compare the interplay between structure and dynamics of spreading for the intimate and ambient networks. In order be able to compare directly, we create a *sampled ambient network*, which contains the same fraction of interactions as the short-range network, but chosen at random among all interactions (Figure 1a). As we show below, the sampled ambient network thus contains both close and distant interactions and shares all salient topological properties with the full ambient network.

**Link weight distributions.** For each of the three networks (ambient, sampled ambient, and intimate), we create a binary-valued adjacency matrix $A_{i \times j \times t}$ with $a_{ijt} = 1$ when an interaction is present and $a_{ijt} = 0$ otherwise; timebins $t$ contain interactions aggregated over $5$ minute intervals. The weight of a link $w_{ij}$ is defined as the total number of interactions occurring on that link $w_{ij} = \sum_t a_{ijt}$. Since the sampled ambient network is generated by sampling interactions at random from the full network, we can calculate the weight distribution for this network analytically (see SI). The distribution of link-weights in all three networks is broad with many weak links (containing few interactions) and a small group of links with very high link-weight (Figure 1b). These broad distributions of weights capture a wide range of social relationships from randomly co-located strangers to pairs of individuals that spend hours together every day.

We structure the comparison between the intimate and ambient networks according to link-weight. First let us consider the weak links. In terms of the weakest links, random sampling of interactions in the ambient networks removes $(1 - f) = 81.7\%$ of all links with weight one. Thresholding by physical distance, however, removes an even greater fraction of low-weight links, with the intimate network retaining only approximately half as many links with weight one as the sampled ambient network. High-weight links are relatively unaffected by removing interactions according to physical distance, so in the intimate network, the highest-weight links maintain $\sim 80\%$ of their interactions. This is in stark contrast to the sampled ambient network, where link-weight is depleted in proportion to the sampling fraction, and high-weight links maintain only $\sim 18\%$ of the interactions from the full ambient network. Overall, the sampled ambient network retains almost twice as many links (26 405, $\sim 62\%$) as the intimate network (13 474, $\sim 31\%$), out of the $42\,838$ links in the full ambient network. Thus, the weight distribution in the intimate network suggests that friends (with high-weight links) tend to be physically close and that most low-weight links correspond to random encounters (encounters between strangers), consistent with results from sociology[17] and network science[15].

**Node entropy is reduced in the short-range network.** From the single node perspective, we find important differences between the intimate and the ambient networks. We can quantify this difference using the Shannon entropy. For a node $i$, we think of each link $w_{ij}$ as a state and and define $\pi(w_{ij}) = w_{ij}/\sum_k w_{ik}$ as the fraction of the node's total contacts taking place on that link. Now, we can define the entropy as $S(i) = -\sum_j \pi(w_{ij}) \log_2 \pi(w_{ij})$. Since infection probability is roughly proportional to link weight (see SI), this quantity can be interpreted as the expected number of yes/no questions needed to establish which of $i$'s connections caused an infection. The distribution of entropy for all three networks is plotted in Figure 2a. For the intimate network (blue), the distribution peaks at $4$ bits, corresponding to an effective group of $2^4 = 16$ potential sources of infection. Due to random sampling on interactions, the distribution of entropies are very similar for the ambient (green) and sampled ambient (orange) networks, emphasizing the structural similarity between these two networks. Both peak at around $6$ bits, corresponding to a much larger group of $2^6 = 64$ potential sources of infection in this network. These results show—for the first time at scale—how the close proximity zone is preferentially reserved for strong ties (e.g. friends or acquaintances) while the distant zone is a more public space where many more random interactions happen, in agreement with the personal zones theory described in the literature[17].

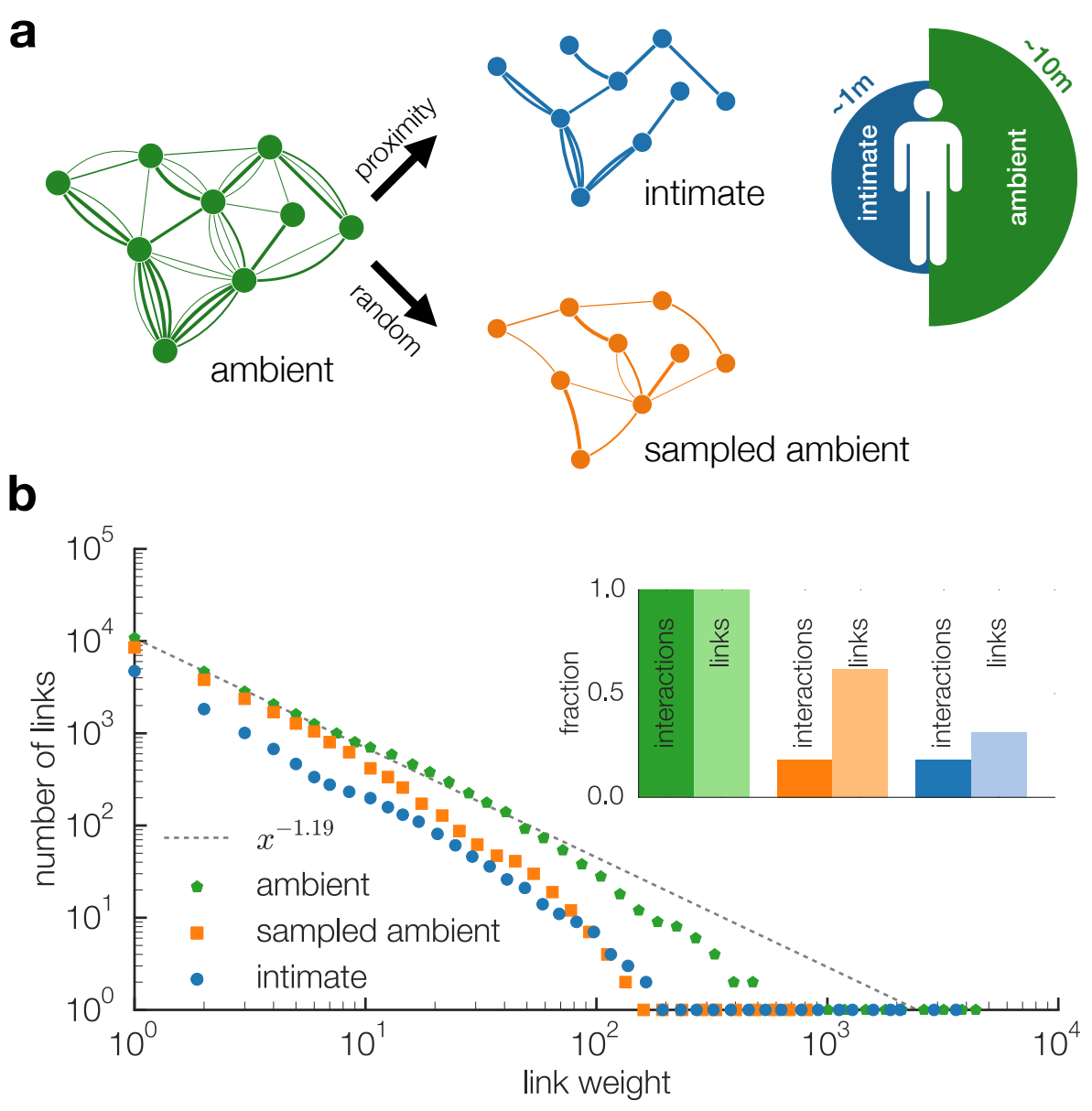

Figure 1: **The network of close proximity interactions. (a)** The full network contains interactions with signal strength used as a proxy for a proximity. In this illustration the dynamic network is integrated over time; edges represent single interactions between participants, and line-width indicates physical proximity. From this full network, corresponding to all edges that support full-range transmission, we create the intimate network by only considering interactions with $\gamma_{ijt} \geq -75dBm$. The sampled ambient network contains the same number of interactions but chosen at random. **(b)** The link weights $(i, j)$ are broadly distributed. The dashed line is a power-law $p(x) \sim x^{-\alpha}$ with $\alpha = 1.19$ inserted as a guide to the eye. The sampled ambient network (orange) has the same number of interactions $\gamma_{ijt}$ as the intimate network (blue), but maintains 62% of links, compared to only 31% of links remaining in the intimate network (inset).

**Distinct strong-to-weak percolation behavior** We now study the interplay between meso-level network structure and tie strength in the intimate and ambient networks. Above we showed that, in the intimate network, a large fraction of interactions takes place on high-weight links. In Figure 2b we quantify how this local effect manifests on the meso- and global scale, by studying the growth of the Largest Connected Component (LCC), as we build the network by adding interactions starting from the strongest links. Specifically, we plot the size of LCC relative to the number of nodes in the network. Note that we study the size of LCC as a function of the fraction of total interactions added, because the three networks do not contain the same number of links. Again, the full and sampled ambient network display identical behavior, with a large connected component starting to form after approximately $20\%$ of interactions have been added to the network. This is in contrast to the intimate network, where the percolation transition does not occur until around $45\%$ of interactions are added to the network (see SI for full percolation analysis).

In both types of networks, the strongest links in the network create small isolated neighborhoods of highly interacting nodes. However, in the ambient networks these communities become connected by adding only a small fraction of connections, while in the short-range network at $50\%$ of interactions the LCC only contains around $25\%$ of the nodes. This structure of highly-connected neighborhoods bridged by weak ties is consistent with structures found e.g. in mobile phone networks[13].

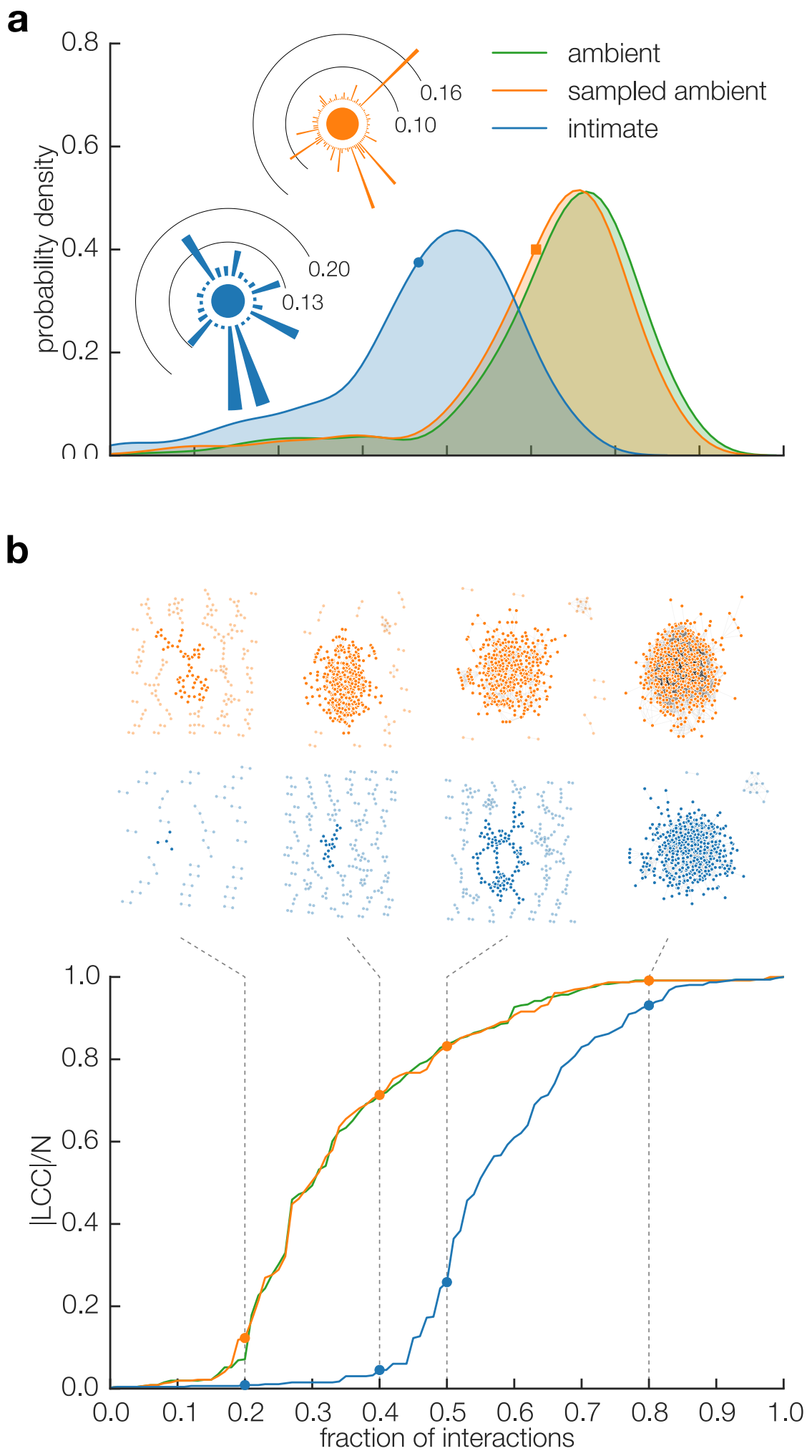

Figure 2: **Difference in network structure.** **(a)** Entropy of contacts. For every node $i$ in the network we calculate Shannon entropy $S(i)$. For a node, the entropy quantifies how many of yes/no questions are needed to identify the source of its infection. Insets show the structure of link weights for a single node (entropy $3.6$) in the droplet network (blue) and in the reduced airborne network (orange, entropy $4.9$), as indicated by the markers on the distributions. Note the similarity between distribution of entropies for the ambient and reduced ambient networks. **(b)** Starting with the strongest links, we add links so they add up to given percentage of interactions in the network (horizontal axis). Keeping track of size of the Largest Connected Component $|LCC|$ relative to the total number of nodes $N$ (vertical axis), we notice that the percolation transition happens at $\sim 20\%$ of interactions in the ambient (green), as well as and reduced ambient network (orange), whereas at the intimate network (blue) percolation takes place at $\sim 45\%$ of interactions. In the intimate network, the pronounced neighborhoods structure contains $45\%$ of all interactions.

**Spreading in short-range network is captured in neighborhoods.** Using a simple Susceptible-Infected-Recovered (SIR) model, we run many simulations of disease spread across the networks, with parameter values chosen to make large outbreaks likely but not guaranteed (see Methods). While we report results for a specific choice of parameters, our results are robust across a wide range values of the transmission parameters. In the intimate network, the simulated pathogen tends to be trapped within small neighborhoods of highly interacting individuals. For each infection event, occurring on link $w_{ij}$, in which node $j$ gets infected, we measure which fraction $I_j$ of the node's direct (1-hop) neighborhood has already been infected. Since this is a weighted network, we define $I_j = W_{\{-i\}}^{-1} \sum_{k \in \mathcal{I}(j), k \neq i} w_{jk}$, where $\mathcal{I}(j)$ is the set of $j$'s infected neighbors and $W_{\{-i\}} = \sum_{k \neq i} w_{jk}$ is the sum of all weights excluding the infecting link. A value of $0$ indicates that no-one in the direct neighborhood besides the infecting node has been yet infected; a value of $0.5$ indicates that neighbors accounting for $50\%$ of link strengths of the nodes have already been infected. Figure 3a, upper panels, shows $I$ for a single epidemic realization in the intimate (left), sampled ambient (middle), and ambient (right) networks; the corresponding lower panels show a kernel density estimation based on $500$ runs of the spreading process.

In the case of the intimate network, we observe behavior consistent with the spreading agent being 'trapped' within neighborhoods, occasionally jumping to a new neighborhood. Early in the epidemic outbreak, when the fraction of infected nodes is low, the disease agent saturates small neighborhoods, and infects new nodes in neighborhoods, where a large fraction ($I > .80$) of neighbors are already infected. Conversely, it is still possible to find neighborhoods with a low fraction ($I < .20$) of infected nodes very late in the epidemic. These effects are possible because the spreading agent does not jump easily between neighborhoods of densely connected nodes.

This picture is in contrast to the full and sampled ambient cases. Here the infection progresses smoothly through the network, unencumbered by social structures. In the ambient networks, the neighborhood infection is roughly proportional to the fraction $F$ of the total network infected, suggesting that the spread starts simultaneously within every graph neighborhood. The kernel density estimation based on $500$ runs, shown in Figure 3a, bottom panels, confirms that this pattern is consistent across random starting conditions (seed node and time). The differences in behavior are quantified in Figure 3b. Here we plot the distribution of $R^2$ of a linear model, fitting infection of the neighborhoods $I$ to the progress of the infection (fraction of network infected $F$), calculated for each realization of an epidemic. For the intimate network, the distribution of $R^2$ peaks at around $0.4$. In the ambient networks, the distribution of $R^2$ peaks at around $0.75$, indicating that direct proportion between the global ($F$) and local ($I$) infection level is a better model for these networks.

**Repeated interactions between infected alters slows spread in short-range network.** In Figure 3c we quantify the tendency for the spreading agent to be trapped within local neighborhoods, by studying how frequently links between two already infected nodes are active. We keep track of how often a link $i, j$ that was infectious in given realization of the spreading process is active again between two infected (or infected and recovered) nodes, measuring the number of interaction 'wasted' within network neighborhoods, which have already been infected. Here we observe a clear difference between the intimate and the sampled ambient network. In the sampled ambient network, where the local connection patterns have high entropy, there is only a low level of activity among infected or recovered individuals. The spreading agent quickly reaches the entire network due to a large number of available paths. This behavior is in contrast to the intimate network, where the disease spread is slower, so links between infected nodes present a larger fraction of interaction events. Eventually, the spreading agent reaches the entire network, transitioning between neighborhoods through close-range interactions on weak/non-existent social ties, e.g. a stranger on the bus. Thus, given the same number of interactions, outbreaks are significantly slower and more contained in the intimate network relative to the sampled ambient case (Figure 3d).

**Differences in dynamics lead to different spreading outcomes.** Figure 4 confirms that the structural differences between the intimate and ambient contact networks lead to different spreading outcomes. Firstly, when the outbreaks do happen in the intimate network, they are smaller, in terms of total number of nodes infected (Figure 4a). Moreover, the probability that an outbreak is contained—reaching only a small fraction of the network ($< 20\%$)—is higher in the intimate network than in the ambient networks (Figure 4a inset). Finally, the time an infection needs to reach $50\%$ of the intimate network is also significantly longer, with the peak of the distribution for sampled ambient network occurring after $7$ days, while the intimate network the peak is delayed to $10$ days (values for a particular choice of transmission and recovery parameters), in spite of the two networks having the same total number of interactions (Figure 4b). In summary, intimate interactions that support droplet transmission are organized in a way that inhibits spreading relative to the ambient case. The sampled ambient network features precisely the same number of interactions as the intimate network, but is structurally identical to the full ambient network according to any network measure. Thus, our results show that taking into account the physical distance of interactions significantly alters the outcome of a simulated outbreak. The qualitative behavior described above is reproduced across a wide range of parameter values.

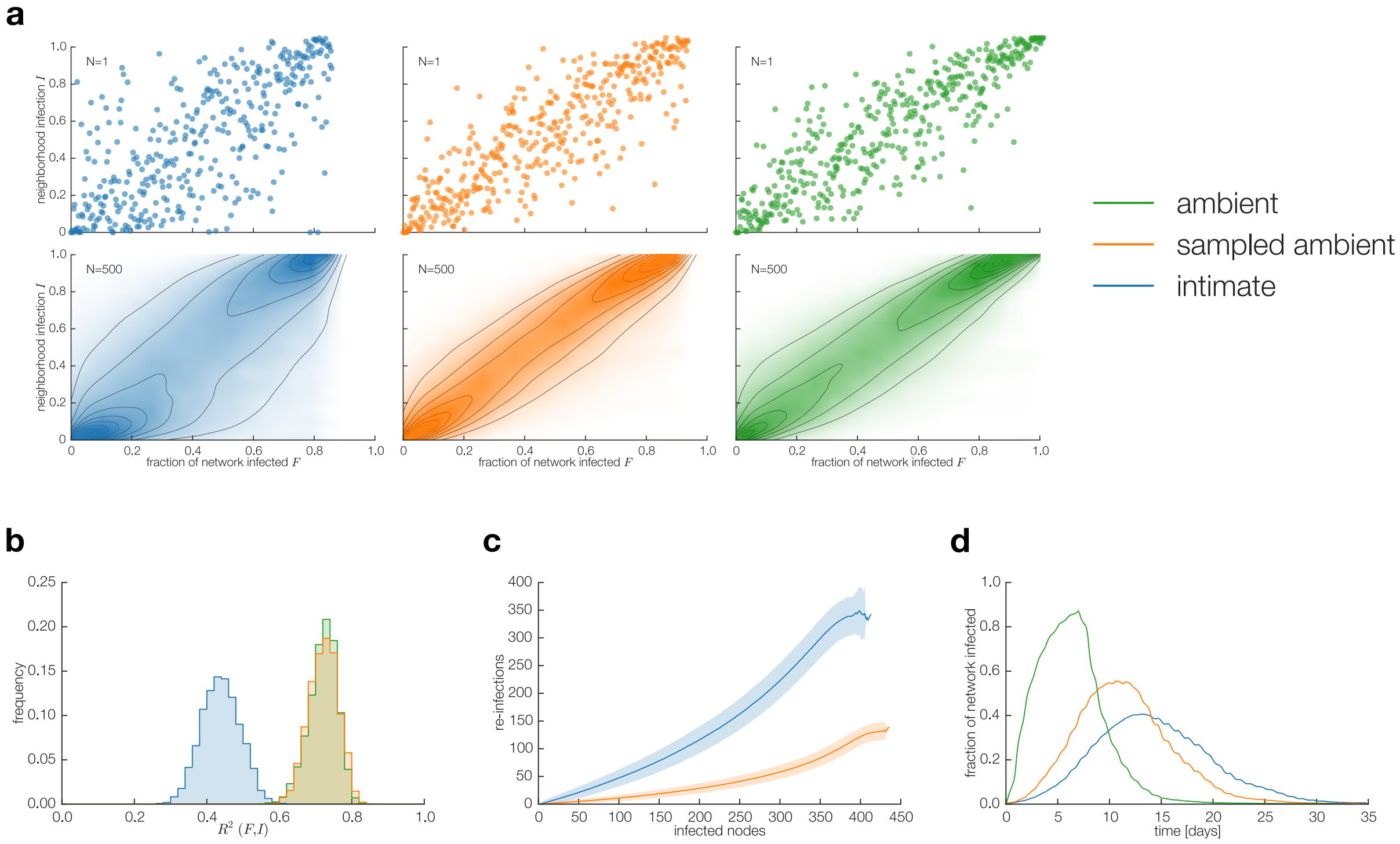

Figure 3: **Dynamics of the spreading process.** Results for 10 000 SIR simulations. **(a)** For every infection event we keep track of the strength of links between the node becoming infected and its neighbors already being infected. In the ambient networks, the infection progresses without major jumps, whereas in the intimate network the simulated pathogen can encounter uninfected neighborhoods even late in the spreading process. **(b)** We quantify the jumps by plotting distribution of $R^2$ between progress of the infection (fraction of network infected $F$) and infection of the neighborhoods $I$. In this sense, time explains $\sim 75\%$ of the variance in the neighborhoods infections in the ambient networks (smooth progress of infection through the network), while only $\sim 40\%$ in the intimate network. **(c)** As the infection progresses through the network, we keep track of how often the a link between two infected nodes is activated. This way we can quantify the 'wasted' infectivity, where the infected nodes connect repeatedly. Shaded areas indicate one standard deviation. **(d)** This results in a slower and more contained outbreak in the intimate network than in the ambient networks.

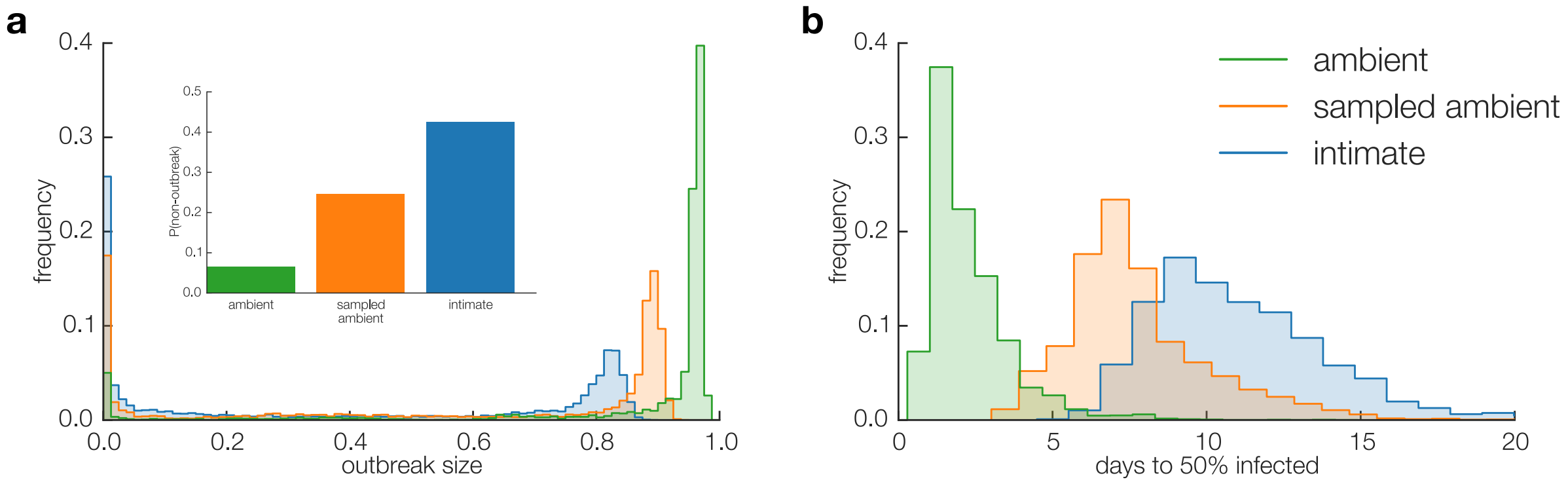

Figure 4: **Statistics of spreading. (a)** In the intimate network the outbreaks are smaller than in the sampled ambient network, even though these two contain exactly the same number of interactions. The probability of outbreak being contained—reaching only a small fraction of the network—is also higher in the intimate network (inset). **(b)** When outbreaks happen, the time to $50\%$ of the network becoming infected is significantly longer in the intimate network, because the spreading is captured within small neighborhoods.

## Discussion

Taking advantage of a large proximity network collected from a densely-connected social system, we have demonstrated a strong structural difference between the intimate networks that support short-range transmission processes and the ambient networks that support transmission across distances up to 10 meters. We did not attempt to create a best possible model of close- and full-range spreading of a real biological agent. The exact process of biological spreading depends on many factors, including events such as sneezing or coughing or pathogen settling on surfaces—physical proximity is just one, although critical, factor to consider. The fundamental differences between proximity networks, however, that we have documented here shows the need for more sophisticated modeling of spreading based on physical proximity of interactions. Specifically, the qualitative differences between spreading across the intimate and ambient networks imply that future measurement and modeling of disease outbreaks must take the mode of transmission into account. In particular, it is critical to consider how data has been collected and not simply categorize any set of co-location observations as a close interactions network.

The proximity of the interactions correlates with tie-strength; we preferentially stay closer to our friends. At the same time, the tie strength is not the same as the distance of interaction; sometimes we are compelled by circumstances to be physically close to a stranger. These close-proximity weak links play a critical role in the spreading process, creating bridges between densely-connected neighborhoods and introducing the spreading to new social circles. The tendency to randomly connect strangers is weak in the droplet network, but in the airborne network the seemingly subtle difference of a few meters introduces many more links between strangers, which allows for faster and more powerful outbreaks to take place, even when considering the exactly same number of interactions. Our findings point to a need for a careful modeling that incorporates the distance of interactions, and highlights a fundamental difference in spreading process between droplet and airborne transmissions.

The fact that proximity of interactions, and thus the mode of transmission, impacts the net-work structure and spreading dynamics is naturally a known fact. For example, networks of sexual contacts are analyzed separately from other types of pathogen spread[5,26], even though both types of networks are physical interactions networks. A central effort in understanding role of physical proximity given by Read et al.[19], where questionnaire data regarding 'close' and 'distant' interactions were collected from 49 participants over 14 non-consecutive days. That study however did not address how differences between the spreading modes would affect the network of infections, with authors calling for larger, more comprehensive studies of differences between the transmission modes. Recently, however, a multitude of new approaches have been developed for collecting data regarding close interactions with the purpose of modeling spreading using various methods, including Bluetooth, RFID, and questionnaires[1,19,27]. Our results imply that in order to be valuable for epidemiological research, new technology driven approaches must draw on domain knowledge and address and incorporate this tacit knowledge about transmission modes.

Spreading processes across proximity networks is highly sensitive to the distance of the interactions. The topology of the droplet network inhibits the spreading process by trapping the infection in tightly-connected neighborhoods, resulting in smaller outbreaks. Thus, regardless of the method of contact tracing—questionnaires, observations, badges, or mobile phones—the recovered networks will differ dramatically depending on the physical proximity of interactions considered.

## Methods

**The dataset** The dataset used in this paper comes from the Copenhagen Networks Study[14]. We use one month of full-resolution data (February 2014). Out of 696 freshmen student participants active in that month we chose students with at least 60% of Bluetooth observations present (resulting median 80%) and who belong to a single connected component. Observations are defined as 5-minute bins in which the user has performed scans, whether the scans contained any devices or not. Since Bluetooth scans do not result in false positives, we symmetrized the observation matrix (resulting in an undirected network), assuming that $\gamma_{ijt} \iff \gamma_{jit}$. This results in improved data quality, with median quality of 85% of 5-minute bins covered. The data is of high temporal resolution (5-minute bins) and provides large coverage of population (out of $\sim 1\,000$ freshman students in total) and time (85% of time).

**RSSI and interaction distance** The received Signal Strength Indicator (RSSI) can be used to estimate the distance between wireless devices[28]. Sekara & Lehmann[15] showed the stability of RSSI in modern mobile phones; the same phones were used in the Copenhagen Networks Study. Based on their results, we use $\gamma_{ijt} = RSSI \geq -75\ dBm$ as an indicator that an interaction was closer than 1 meter. This value can be considered a conservative estimation, as the measurements in Ref.[15] have been performed without obstacles. Thus, we expect that $\gamma_{ijt} \geq -75\ dBm$ may not include all the close interactions, but it will not include distant interactions. When the interaction matrix is symmetrized, we take the smallest distance (largest RSSI) that happened between users in given timebin $\gamma_{ijt} == \gamma_{jit} = \min(\gamma_{ijt}, \gamma_{jit})$.

The approach presented here has some limitations. While all mobile phones used for data collection in the study were all the same model and the obtained RSSI values are comparable, it is important to emphasize that our distance threshold is noisy; RSSI may differ depending on where the phone is placed, environmental conditions, etc. In that sense, our results can be considered a *lower bound* of the difference between the two types of networks, since a perfectly noisy threshold would produce two randomly sampled networks with no difference between them.

**Epidemic simulations** To show the dynamics of the spreading process in the droplet and airborne networks we use a simple Susceptible-Infected-Recovered (SIR) simulation. We run a large number of simulations ($N = 10\,000$) on the full temporal network, where every interaction between Infected and Susceptible participants can lead to infection with probability $\beta = 0.02$. Users stay in infected status for $\mu_t = 7$ days, after which they are moved to Recovered state and cannot be re-infected. The starting timebin and seed node are chosen at random in every simulation and used for simulation on all three networks (ambient, sampled ambient, and intimate). We use one month of data (28 days, 8 064 5-minute timebins) with periodic boundary conditions. The parameter values are chosen so that outbreaks are likely, but not guaranteed and with sizes that do not trivially saturate the entire network. The qualitatively behavior of our analysis is unchanged across a wide range of parameter values.

**Acknowledgements** We thank Piotr Sapiezynski, Vedran Sekara, and Yves-Alexandre de Montjoye for useful discussions. This research was funded, in part, by the Villum Foundation ("High Resolution Networks"), as well as the University of Copenhagen through the UCPH-2016 "Social Fabric" grant.

**Competing Interests** The authors declare that they have no competing financial interests.

**Correspondence** Correspondence and requests for materials should be addressed to Arkadiusz Stopczynski (email: arks@dtu.dk).

# Supplementary Information for Physical Proximity and Spreading in Dynamic Social Networks

Arkadiusz Stopczynski[1,2], Alex 'Sandy' Pentland[2], Sune Lehmann[1,3]

[1] *Department of Applied Mathematics and Computer Science, Technical University of Denmark, Kgs. Lyngby, Denmark*

[2] *Media Lab, Massachusetts Institute of Technology, Cambridge, MA, USA*

[3] *The Niels Bohr Institute, University of Copenhagen, Copenhagen, Denmark*

## 1 The dataset

For every participant in the Copenhagen Networks Study [1] in February 2014 (696 active users) we calculate the number of 5-minute timebins in which we have data on Bluetooth obsevations (whether they contain any other participant in the study or not). Out of this population we choose users with data in at least 60% of timebins covered (Figure 5). As the Bluetooth scans do not yield false positives (devices that are not actually present during the scan), we make the discovery network symmetric by assuming that if user $i$ observed user $j$ in timebin $t$ the opposite is also true. After the matrix is made symmetric, the median time coverage for the users reaches 84%. The degree and node strength have broad distribution in the full network (Figure 6). The strength of the nodes has a slightly longer tail in the intimate network, indicating existence of a few nodes with very strong links that are reduced in the random sampling.

The temporal dynamics of the social networks show an expected daily and weekly pattern as quantified by the number of active links in the network in 1-hour timebins (Figure 7). Weekdays display morning and afternoon peaks of activity, with decreased activity during lunchtime. Fridays show a significantly reduced afternoon peak but a much more pronounced evening peak, indicating social gatherings. Most of the weekends show significantly smaller activity, except for occasional parties (for example on Saturday evening in day 22). This shows that although the analyzed network is primarily driven by work-related interactions, it also contains a non-trivial number of social interactions.

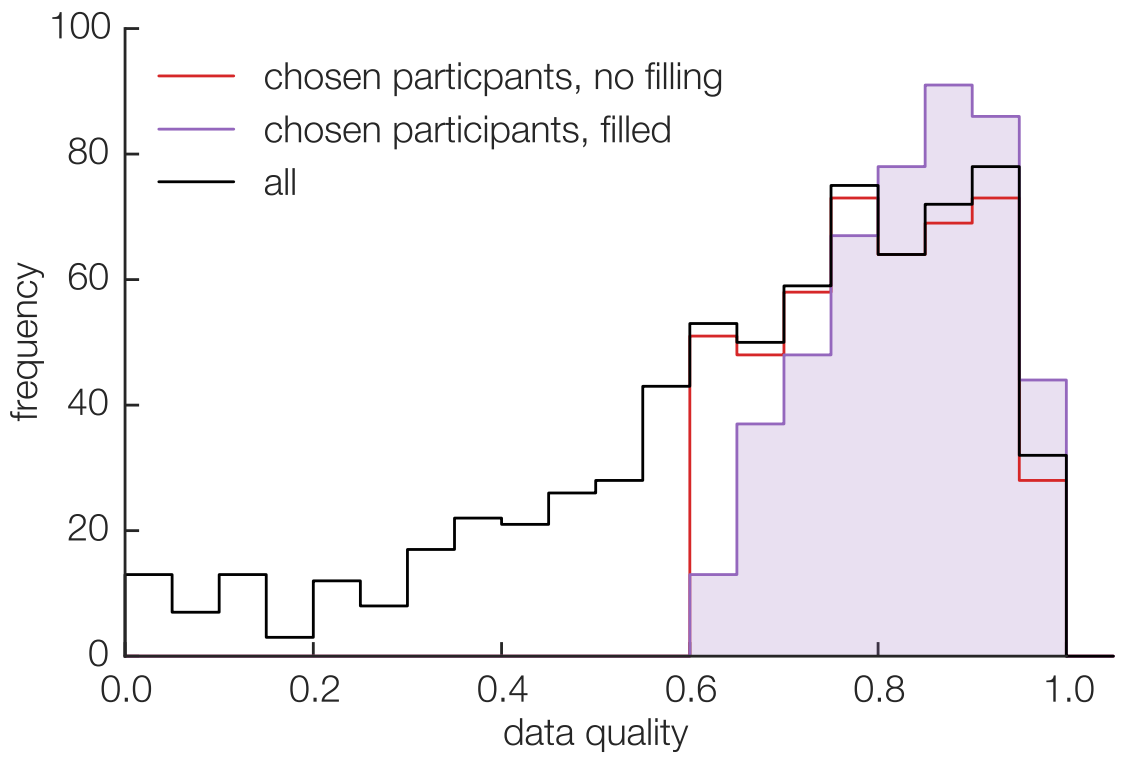

Figure 5: **Quality of the data.** We calculate in how many 5-minute bins we have Bluetooth data for the users. Out of all users we chose these with data quality of over 60%. After filling the observations by making the observations matrix symmetric, the data quality still improves slightly (shaded area).

The difference in the entropy of interactions between intimate and (sampled) ambient network (shown in the main text) is not explained by a simple change in the degree of the nodes. Plotting the difference in entropy vs. change in degree reveals only week correlation $R^2 = 0.17$ as shown in Figure 8.

**Link weights and spreading** Strong links in the networks also tend to be more infectious. This is shown in Figure 9. The correlation is strongest in the sampled ambient network ($R^2 = 0.90$). In the intimate (and ambient) network the infectious power of the strongest links is lower than we would expect from a linear model, due to the high clustering of these links in the tightly-connected neighborhoods (as shown in the main text).

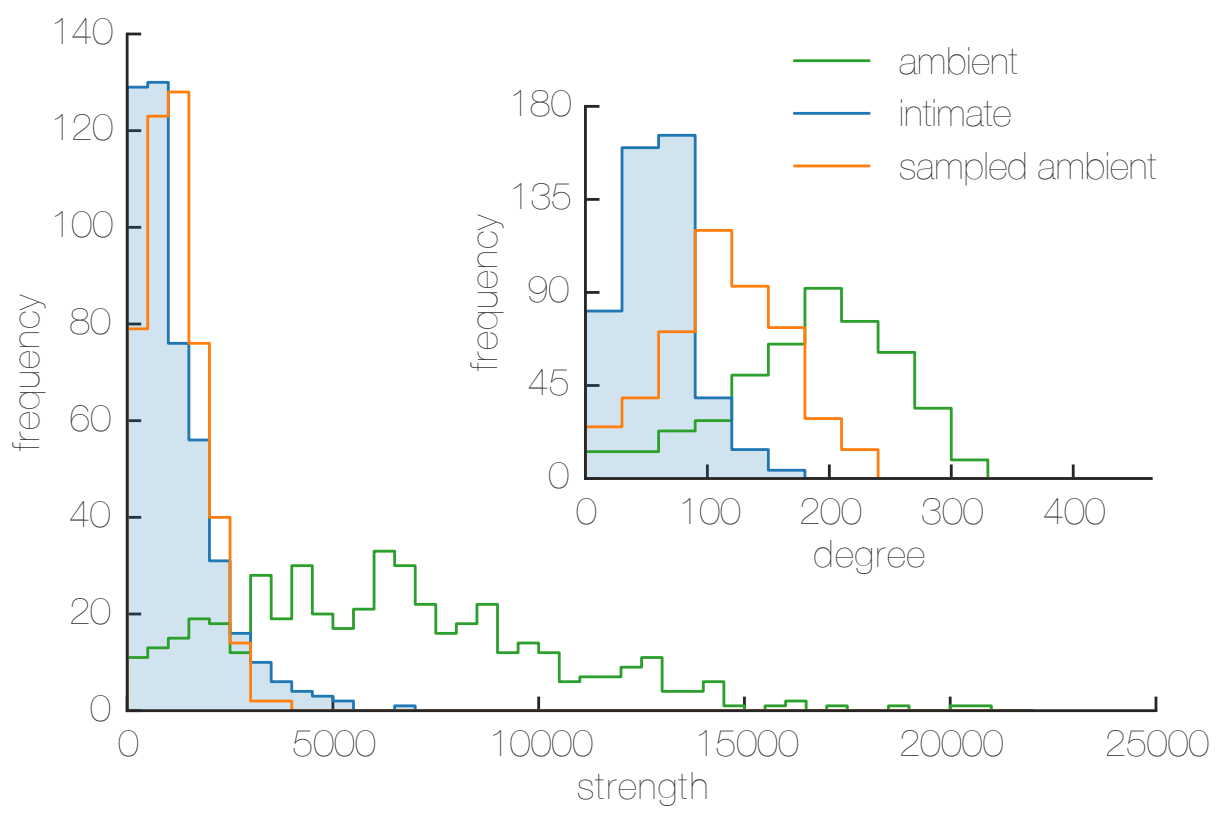

Figure 6: **Degree and strength in the network.**

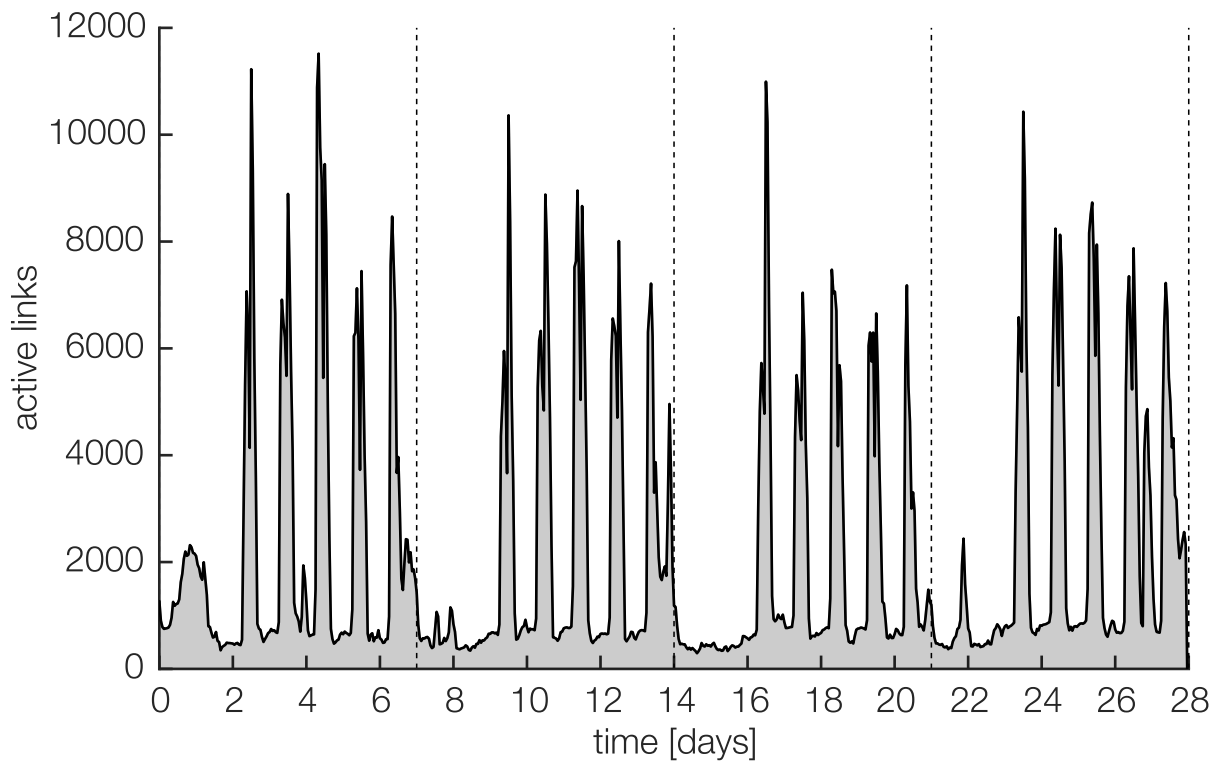

Figure 7: **Temporal dynamics in the network.** Number of active links in 1-hour temporal bins. Dotted lines show Friday-Saturday midnight.

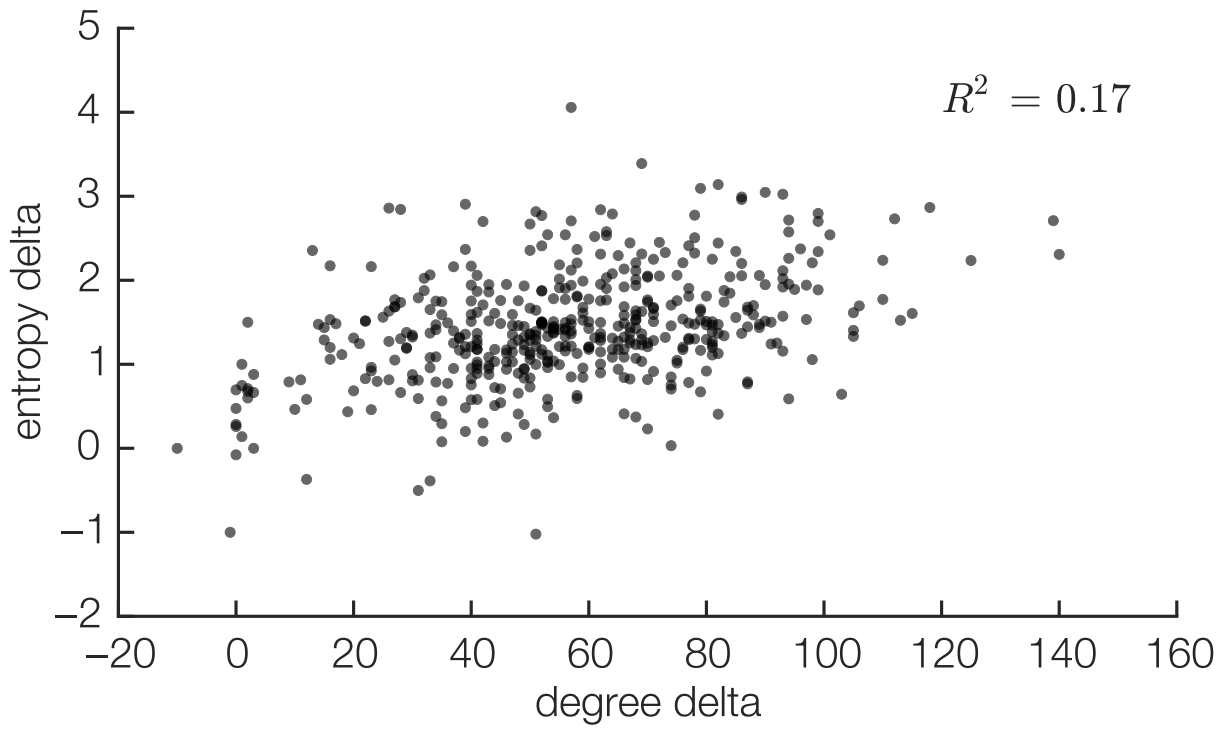

Figure 8: **Correlation between difference in degree and entropy between intimate and sampled ambient networks.** The change in the degree does not explain the change in the entropy.

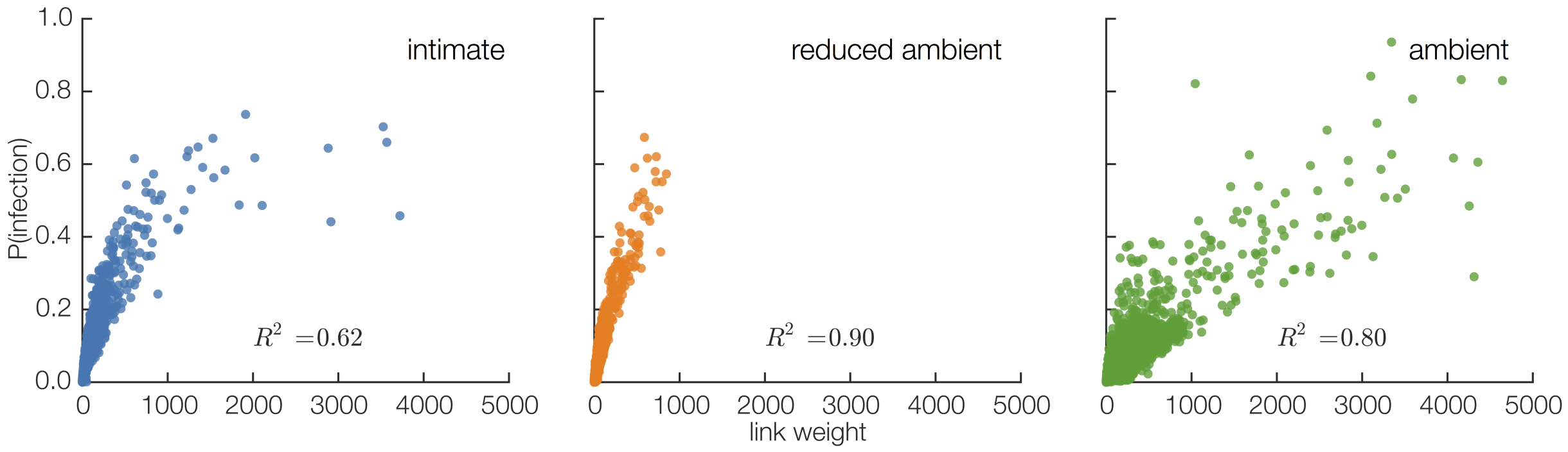

Figure 9: **Correlation between link weight and probability of infection happening on this link (in any direction).**

## 2 RSSI and interaction proximity

RSSI is a noisy proxy for distance. Numerous lab experiments have shown strong (logarithmic) relation between RSSI and distance [2–4]. Recently, the relation between proximity measured by RSSI and social ties has also been shown [5]. The distribution of RSSI observations decays exponentially (Figure 10). The dataset used here does not contain data allowing for direct validation of RSSI as proximity indicator, as it does not include any signal that can reliably identify proximity with higher or even comparable resolution. Instead, we use the overlap of other sensed Bluetooth devices to show how RSSI relates to physical proximity.

For every observations between participants we calculate the jaccard overlap between all the other devices they sensed in the timebin. For example, for a 10 minute timebin—which on average contains two scans—we take the value of RSSI as the smallest RSSI between the users in that timebin and for every user we create a set of all unique devices sensed in that period (usually in two scans). Distant interactions (with small RSSI) show smaller overlap of the discovered devices (Figure 11). This is true for timebins (over which the overlap is calculated) spanning over four orders of magnitude, including extremely short windows of 5 seconds and long windows of 4 hours. In the timebins shorter than used Bluetooth scanning interval (5 minutes) the overlap between single scans is calculated (and the scans can be separated in time up to the size of the timebin). For the longer timebins, overlap between devices sensed in multiple scans is used. For the high RSSI values there is a negligible fraction of observations that do not show at least 50% overlap (distributions in Figure 11), which confirms that higher RSSI values are indicative of closer physical proximity.

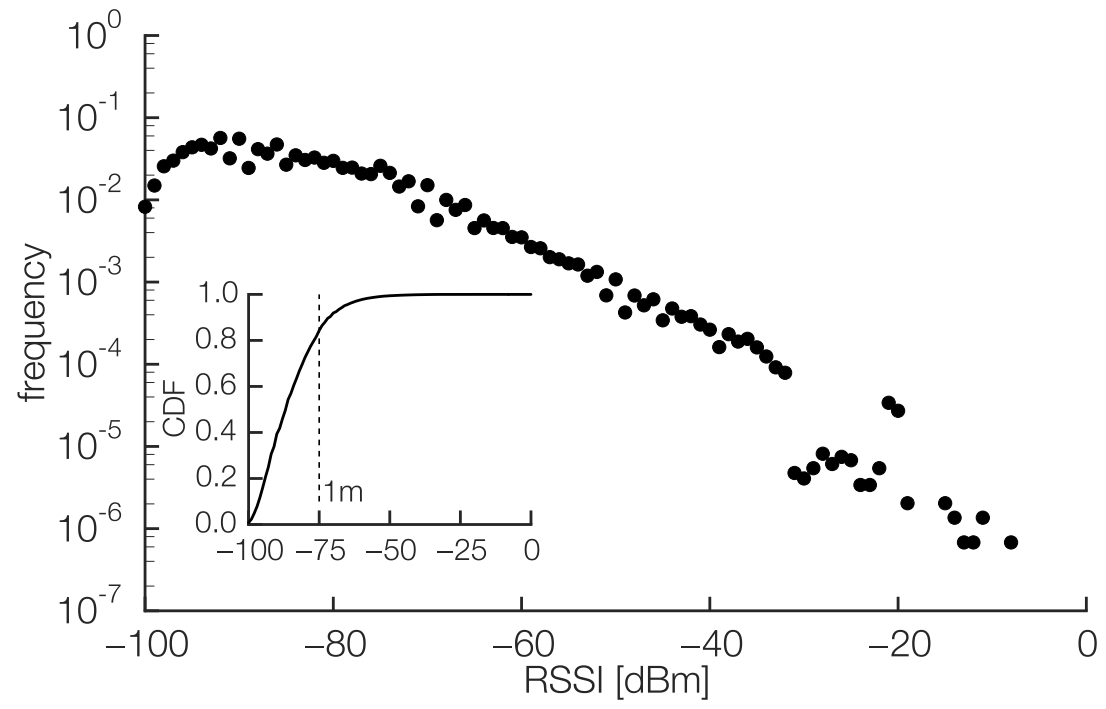

Figure 10: **Distribution of RSSI for the full network.** Short-range interactions (1m and closer) account for 18.3% of all interactions (inset).

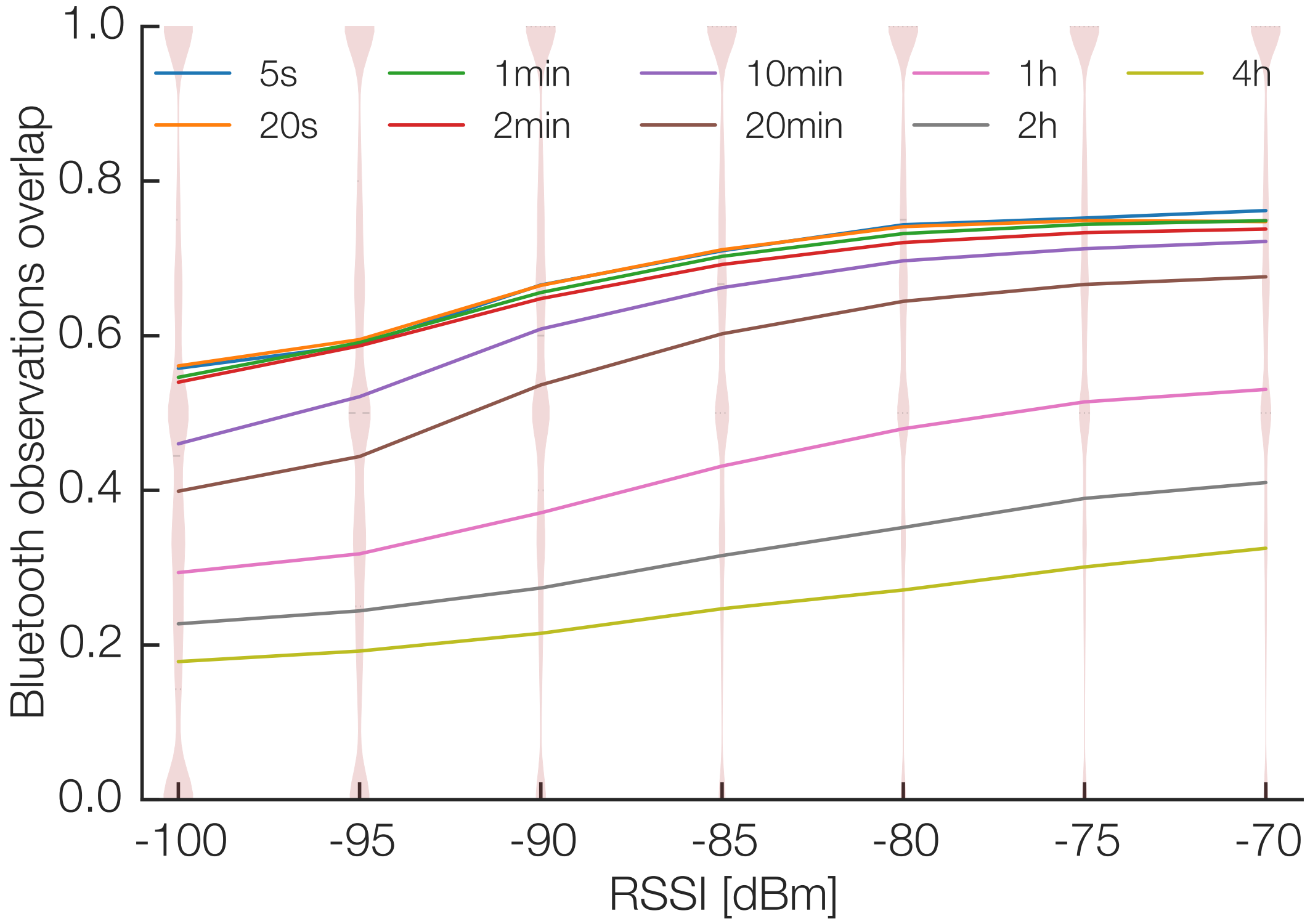

Figure 11: **Jaccard overlap between sensed Bluetooth devices depending on the RSSI between participants.** Higher RSSI indicates closer physical proximity which also results in the higher overlap. Observation are binned in $5dBm$ bins. Plots for different time windows, across which the overlap is calculated. For all the time windows the higher RSSI indicates higher overlap of discovered devices. Lines indicate mean values. Distributions are shown for $t = 10$ min).

## 3 Position of the phase transition point

In order to determine the at which fraction of interactions the Largest Connected Component (LCC) starts forming as shown in Figure 2 (main text), we carry the finite size scaling (FSS), analogous to the one described by Onnela et al. [6]. As we show in the main text, adding links accounting for increasing fraction of interactions leads to a sudden formation of the LCC. The point where we observe the phase transition $f_c$ is different for intimate and ambient networks (both full and sampled). This difference in the transition points indicates difference in the structure of the networks, where in the intimate network strong links create disconnected neighborhoods. As the system size is relatively small (N=464), we are interested whether and to what extent the effect we observe is caused by the finite size of the system or does it indicate a real structural difference.

We subsample all three networks, choosing nodes at random with probability $p \leq 1$, keeping only links between chosen nodes. We adjust $p$ to start with system size of $\approx 110$ and move up in steps of 10. As such random sampling can lead to network becoming fragmented, we only take the largest connected component of the subsampled networks. We then build every subsampled network starting with the strongest links and keep track of the size of the LCC in order to identify the point of phase transition at which the LCC starts growing rapidly (same procedure as for the full network in the main manuscript in Figure 2).

For every network we identify the point of the phase transition $f_c(N)$. The point may be challenging to find exactly, especially in the smaller systems as the size of LCC, although monotonous, tends to show multiple jumps. We employ two simple methods to identify the point. The size of LCC is expected to vary most rapidly at the threshold, so we define the transition point as the maximum of $\partial R_{LCC}/\partial f$. The second approach is to define a fixed threshold at the size of LCC, high enough that it is rarely reached before the jump occurs and small enough that it is sensitive to the jump point. We estimate from the data the point at size of LCC containing $15\%$ of the all nodes in the network. The results are shown in Figure 12 (intimate) and Figure 13 (sampled ambient). We find that threshold-based inference of the transition point performs better than a derivate-based one, with points identified close to what we naturally perceive as a threshold. In the further analysis we use the threshold-based method of transition point identification.

Plotting the identified transition points against $1/N$, we fit second order polynomial to calculate $f_c(\infty)$ as $y$-intercept point, i.e. critical threshold at $N \to \infty$. The plots for a single realization of subsampling are shown in Figure 14. We run the subsampling (all the N steps) 1 000 times and plot the distribution of the calculated $f_c(\infty)$, as shown in Figure 15. The values are normally distributed and the distributions for sampled and full ambient networks overlap. The calculated $f_c(\infty)$ for the intimate network (the mean value of the distribution) is $0.36 \pm 0.05$. The value for both full and reduced airborne network is $0.08 \pm 0.04$. Value of $f_c(\infty) = 0$ would indicate no phase transition, with LCC starting to grow as soon as the network is constructed. Since both values are significantly different from 0 we conclude that we observed a genuine phase transition for both intimate and ambient networks. However, for the intimate network the $f_c(\infty)$ is much higher, consistent with the finding reported in the main text for $f_c(464)$.

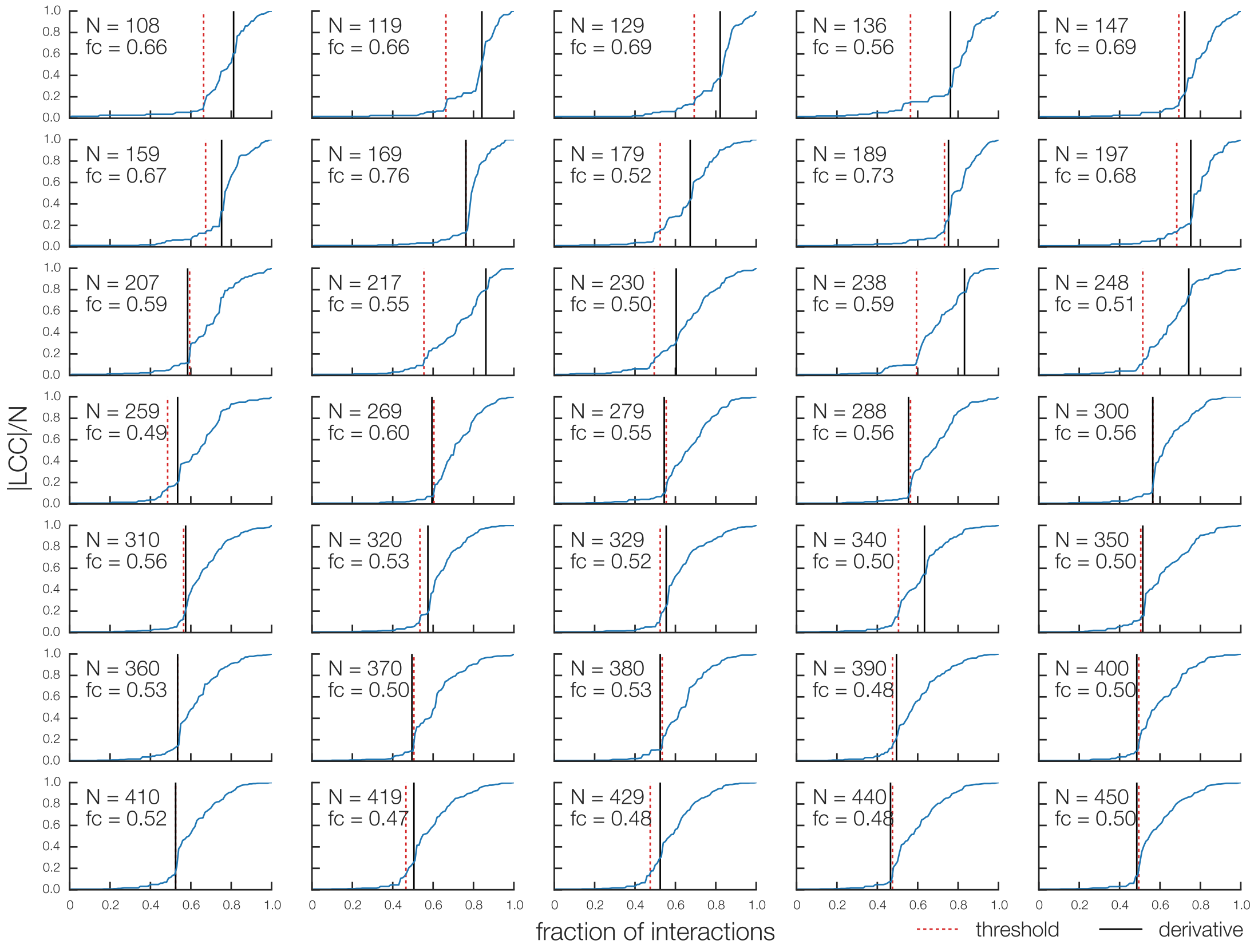

Figure 12: **Transition point identification for intimate network.**

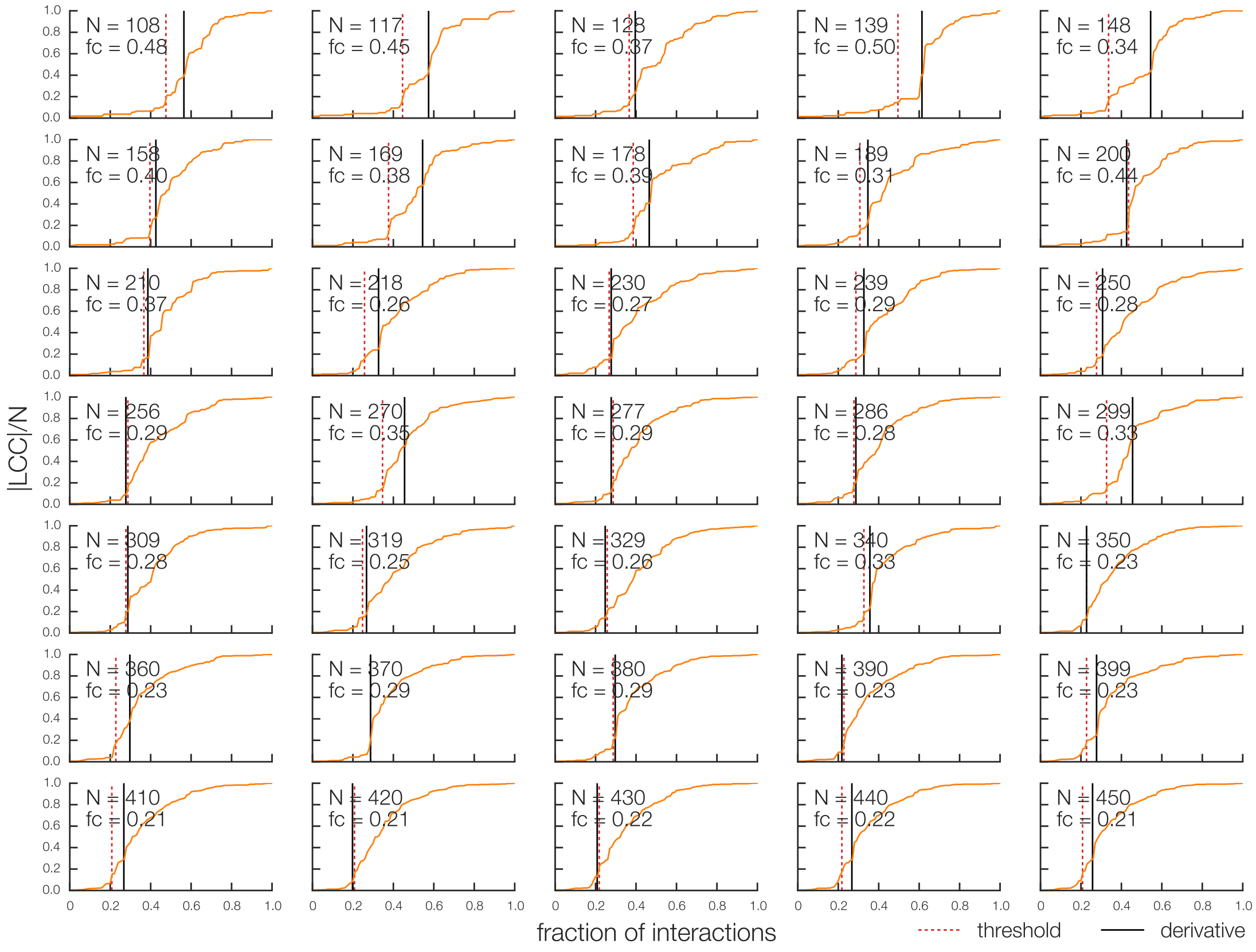

Figure 13: **Transition point identification for sampled ambient network.**

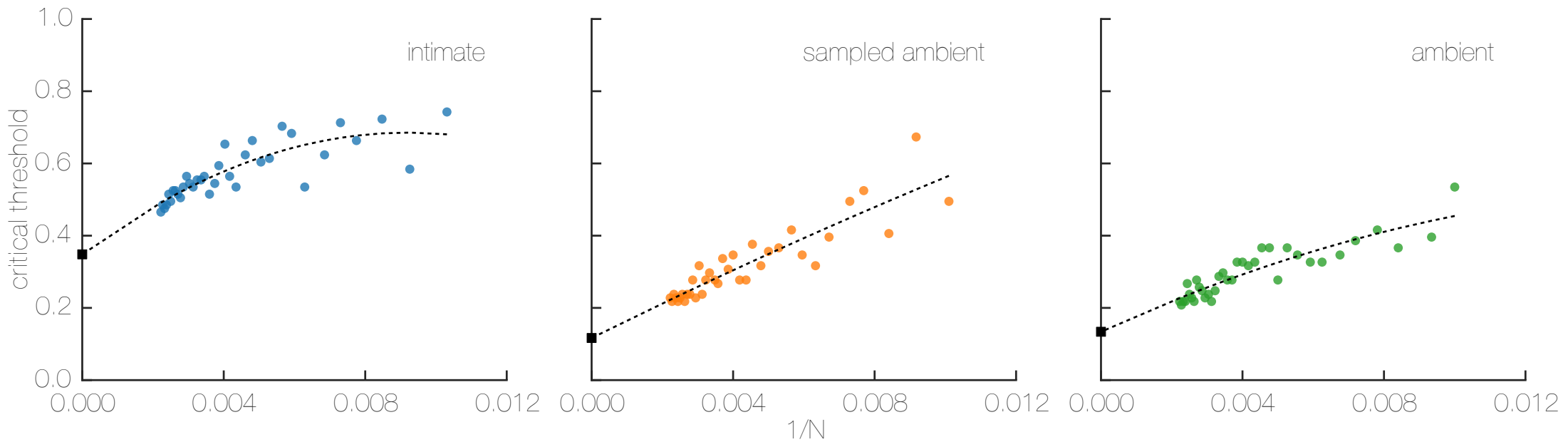

Figure 14: **Critical threshold $f_c(N)$ plotted against $1/N$.** Second order polynomial is fitted to the data and $y$-intercept indicates the value of $f_c(\infty)$.

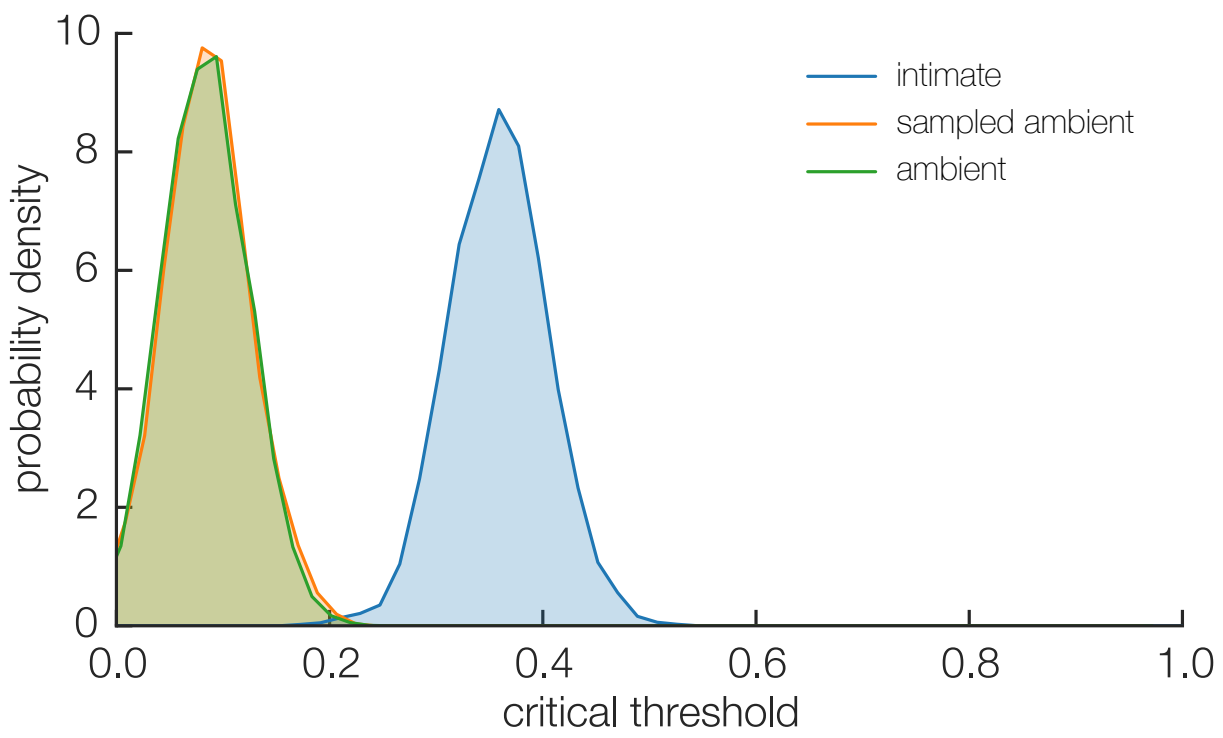

Figure 15: **Distribution of calculated $f_c(\infty)$.** As all distributions are significantly higher than 0 we conclude that we observe a genuine $f_c(464)$ for all the networks.